\documentclass[aps,prl,showpacs,twocolumn,superscriptaddress]{revtex4}
\usepackage{graphics}
\usepackage[dvips]{color}
\usepackage[T1]{fontenc}
\usepackage[latin1]{inputenc}
\usepackage{graphicx}
\usepackage{amssymb}
\usepackage{rotating}
\usepackage{epsfig}

\input epsf.sty

\begin{document}

\title{
Kinetically driven helix formation during the homopolymer collapse process
}
\author{Sid Ahmed Sabeur}
\author{Fatima Hamdache}
\affiliation{
D\'epartement de Physique, Facult\'e des Sciences, USTO,
Oran 31000, Algeria
}
\author{Friederike Schmid}
\affiliation{
Fakult\"at f\"ur Physik, Universit\"at Bielefeld, D -- 33615 Bielefeld,
Germany
}
\date{\today}
\begin{abstract}

Using Langevin simulations, we find that simple 'generic' bead-and-spring
homopolymer chains in a sufficiently bad solvent spontaneously develop helical
order during the process of collapsing from an initially stretched
conformation.  The helix formation is initiated by the unstable modes of
the straight chain, which drive the system towards a long-lived metastable
transient state. The effect is most pronounced if hydrodynamic interactions
are screened.

\end{abstract}


\maketitle


Helices are among the most common structures in nature.
They are simple shapes which can be stabilized rather easily with
suitable potentials~\cite{kemp1,rapaport,banavar}.  $\alpha$-helices
in proteins have the favorable property of being highly
``designable'' in the sense that they are robust to mutations
and thermodynamically stable~\cite{miller}. Hence helices are
interesting structures from a static point of view. In this Rapid Communication,
we report on a dynamic nonequilibrium phenomenon where they
also appear spontaneously, for purely kinetic reasons, without
special potentials or local constraints~\cite{maritan}:
In computer simulations, we find that simple, ``generic'' polymers in
suitable environments recoil into well-ordered helices when released
from an initially stretched configuration. This happens despite the fact
that the ground state of the polymers is definitely non-helical.
The helices emerge as metastable transient states with a finite,
temperature-dependent lifetime, which may become quite large.

The dynamics of polymer collapse has been studied in generic model chains
by a number of authors~\cite{dawson,frish,yethiraj,kikuchi}.
Our work differs from these studies in two aspects. First, we study quenches
to very low temperatures, about 1/100 of the Theta temperature. Second,
we do not start from an equilibrated high-temperature conformation (a coil),
but from a stretched chain. Hence we study ``mechanical'' quenches rather
than temperature quenches. Such rapid mechanical quenches are realized,
{\em e.g.}, if the chain is attached at both ends to objects that move
apart, by links that are weak enough that they eventually
break up.


We have observed helix formation for various different choices of potentials.
In this Rapid Communication, we restrict ourselves to the simple case of Lennard-Jones
beads connected by harmonic springs for clarity of presentation.
Beads that are not direct neighbors on the chain interact {\em via}
a truncated Lennard-Jones potential
\begin{equation}
\label{eq:vlj}
V_{\mbox{\tiny LJ}}(r) = \left\{
\begin{array}{l}
4 \epsilon [ (\sigma/r)^{12} - (\sigma/r)^6 + c_0 ]
\; \mbox{for $r<2.5 \sigma$}\\
0 \quad \mbox{otherwise},
\end{array}
\right.
\end{equation}
where the constant $c_0$
is chosen such that the potential is continuous everywhere. Adjacent beads
in the chain are connected by bonds subject to the spring potential
\begin{equation}
\label{eq:vbond}
V_{\mbox{\tiny bond}}(r) = a \: (r - r_0)^2
\end{equation}
with $a = 100 \epsilon/\sigma^2$ and $r_0 = 0.85 \sigma$. The theta temperature
for these chains is of the order $k_B T_{\theta} \sim \epsilon$~\cite{taylor}.
We investigate the kinetics of chain collapse at temperatures that are
10-100 times lower, starting from an initially straight chain. As dynamical
model for the motion of the beads we use Langevin dynamics with and without
hydrodynamics. Inertia effects are taken to be negligible~\cite{fn0}.
The natural units of our
simulation are defined in terms of the bead size $\sigma$, the Lennard-Jones
energy $\epsilon$, and the friction coefficient $\zeta$ (see below). Based
on these quantities, the time unit is $\tau = \zeta\sigma^2/\epsilon$.


We begin with discussing the collapse dynamics without hydrodynamics.
The solvent surrounding the chain is then effectively replaced by the
friction $\zeta$ and a Gaussian distributed stochastic force $\vec{\eta}_i$
acting on the monomer $i$, which fulfills the conditions~\cite{doi}
$<\vec{\eta}_i>  =  0$ and
\begin{equation}
  \label{eq:eta}
  <\eta_{i,\alpha}(t) \eta_{j,\beta}(t')>
  = 2 \zeta \: k_B T \: \delta_{ij} \delta_{\alpha\beta} \: \delta(t-t'),
\end{equation}
with monomer indices $i, \; j = 1 \dots N$, cartesian directions
$\alpha,\beta \in \{x,y,z\}$, and $t$, $t'$ two given times. The
equations of motion are
\begin{equation}
  \label{eq:dynamics}
  \zeta \: \dot{\vec{r}}_i = \vec{f}_i + \vec{\eta}_i.
\end{equation}
They are integrated using an Euler algorithm with the time step
$\Delta_t = 5\times 10^{-4} \tau$. The stochastic noise $\vec{\eta}_i$ was
implemented by picking random numbers with uniform distribution, but the
correct mean and variance (\ref{eq:eta}) at every time step~\cite{duenweg}.

\begin{figure}[t]
\includegraphics[scale=0.55,angle=0]{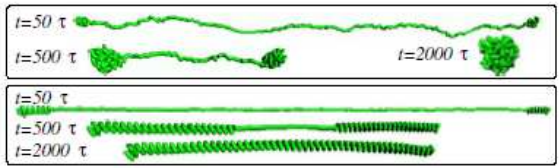}
\vspace*{-0.3cm}
\caption{\label{fig1}
Configurations of an initially stretched chain with length $N=200$ monomers
at different times $t$ for the temperature $T=0.5 \epsilon/k_B$ (top) and
$T=0.01 \epsilon/k_B$ (bottom).
}
\end{figure}

Figure.~\ref{fig1} shows examples of a collapsing chain at moderate and very
low temperature ($T = 0.5 \epsilon/k_B$ and $T=0.01 \epsilon/k_B$, respectively).
At moderate temperature, a variant of the well-known pearl-necklace
scenario~\cite{dawson,frish} is recovered: The collapse starts with the
formation of small globules at both ends of the chain, which subsequently
grow in size and finally merge into a single globule.
At low temperature, the scenario is entirely different:
Instead of disordered globules, well-ordered helices appear at the ends of
the chain.  This process is symmetry-breaking, since the model potentials do
not impose any chirality. The helical structures then propagate along the
chain, until they meet in the middle. The final chain structure is an almost
perfect helix with possibly one defect in the middle, depending on whether
the chiralities of the two merging helices match each other. It is
characterized by the radius $R \sim 0.5 \sigma$, the pitch $D \sim 1 \sigma$,
and has roughly $n\sim 4$ monomers per turn.

Simple energy considerations show that these helices are by no means the
ground state of the chain. The lowest energy state of an infinitely long
chain, where surface effects can be neglected, is most likely a structure
with close-packed parallel strands. A straightforward energy minimization
for this structure yields the energy per monomer
$U_{\mbox{\tiny tot}}/N = -7.75 \epsilon$.
In contrast, the lowest possible energy per monomer of an infinitely long
helix is $U_{tot}/N = -3.26 \epsilon$, which is clearly higher. This is of
course not surprising, given that helices have a large, energetically
unfavorable polymer-solvent interface. They cannot possibly be ground states
at infinite chain length. The question remains whether the finitely long
chains considered in the simulations have a helical ground state.
The exact calculation of the ground state energy for chains of
finite length is a formidable task. We have estimated an upper bound with the
method of simulated annealing. At chain length $N=50$, we get
$U_{\mbox{\tiny tot}} < -210 \epsilon$, corresponding to
$U_{\mbox{\tiny tot}}/N < - 4.2 \epsilon$ per monomer, which is
well below the energy of an optimized helix. The low energy conformations do
not even exhibit local helical order. Hence we conclude that chains of length
$N=50$ or longer are not helical: The helical state observed in the simulations
is a kinetic trap.

Intriguingly, it does not even correspond to the energetically
optimized helix. The energy minimization for helical structures (minimized
with respect to the radius, the pitch, and the bond length) yields a
series of local minima characterized by different numbers $n$ of monomers
per turn, {\em i.e.}, $n=3.92, 4.81, 5.75, 6.71, \cdots$. The lowest minimum
is at $n=4.81$.  The structures of the helices shown in
Fig.~\ref{fig1} are closer to the structure corresponding to $n=3.92$
with an energy per monomer $U_{\mbox{\tiny tot}}/N = -3.04 \epsilon$.

Insight into the initial process of helix formation can be gained from a
linear stability analysis of the starting conformation, the stretched chain.
The unperturbed reference state is an energetically relaxed straight chain
with bond length $b \approx b_0$. The Eigenmodes of the Hessian matrix are
the longitudinal ($L$) and transverse ($T$) phonons of this chain.
Neglecting chain end effects, they have the wavelengths
$\lambda_m = b N/m$ and the frequencies~\cite{fn1}
\begin{eqnarray}
\label{phonons_l}
\omega_{m,L}^2 &=& a + \frac{d^2 V_{\mbox{\tiny LJ}}}{dr^2}
\big(\cos(2 \pi b/\lambda_m) +1\big)\\
\label{phonons_t}
\omega_{m,T}^2 &=& \frac{d V_{\mbox{\tiny LJ}}}{dr} \frac{1}{2 b}
\big(\cos(2 \pi b/\lambda_m) -1\big),
\end{eqnarray}
where $d^2 V_{\mbox{\tiny LJ}}/dr^2$ is negative and $d V_{\mbox{\tiny LJ}}/dr$
is positive. Unstable modes are characterized by imaginary frequencies. If the
spring constant $a$ is sufficiently large, the straight chain conformation is
stable with respect to all longitudinal phonons. However, it is unstable with
respect to transverse phonons with short wavelengths, $\lambda < 4b$. Hence
helical modes with less than $n=4$ beads per turn are unstable.
The most unstable phonon mode corresponds to a zigzag mode.
Indeed, weak zigzags are observed at the onset of the collapse process,
but helix formation soon takes over, due
to the fact that the helical state corresponds to a local energy minimum.
We conclude that the helix formation is initially a driven process.
This explains why the system picks the helical state with
$n=3.92$, rather than the optimized helical state at $n=4.81$.

The subsequent collapse is analyzed most conveniently in a slightly modified
system where one chain end is fixed, corresponding, {\em e.g.}, to a grafted chain.
The helix formation then proceeds from only one end (the free end), and the defect
in the middle of the chain is avoided. Following Kemp and Chen~\cite{kemp2}, we
define the two helical order parameters
\begin{equation}
\label{eq:order}
H_4 = \big( \frac{1}{N-2} \sum_{i=2}^{N-1} \hat{u}_i \big)^2,
\quad
H_2 = \frac{1}{N-3} \sum_{i=2}^{N-2} \hat{u}_i \cdot \hat{u}_{i+1}.
\end{equation}
where the $\hat{u}_i$ are unit vectors proportional to
$\vec{u}_i \propto (\mathbf{r}_i - \mathbf{r}_{i-1}) \times
(\mathbf{r}_{i+1} - \mathbf{r}_i)$, and $N$ is the chain length.
The parameter $H_4$ characterizes the global helical order, and $H_2$ the local
order along the chain.

\begin{figure}[t]
\includegraphics[scale=0.55,angle=0]{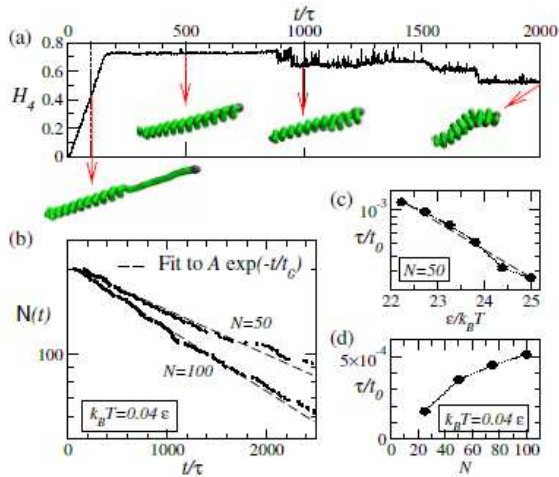}
\vspace*{-0.3cm}
\caption{\label{fig2}
Time evolution of helical order for endgrafted chains with length $N=50$
at temperatures around $k_B T/\epsilon \sim 0.04$.
(a) Evolution of order parameter $H_4$ for one specific run,
with corresponding chain conformations
(grafting point is represented by a dark sphere).
(b) Number N$(t)$ of remaining purely helical conformations after the time $t$
in a sample of 200 initially helical conformations, at temperature
$k_B T/\epsilon = 0.04$ for two chain lengths as indicated. Dashed lines
show fits to an exponential, N$(t) \sim \exp(-t/t_0)$ with escape time $t_0$.
(c,d) Escape rates $1/t_0$ in units of $1/\tau$ as a function of chain
length for the temperature $k_B T/\epsilon = 0.04$ (c), and as a function
of the inverse temperature for the chain length $N=50$ (d). Dashed line
in (c) shows the result of a fit to $1/t_0 \propto \exp(- \Delta E/k_B T)$,
giving $\Delta E = 0.041 \pm 0.002 \epsilon$.
}
\end{figure}

Fig.~\ref{fig2}a) shows a typical time evolution of the order parameter
$H_4$ for such a chain at the temperature $T=0.04 \epsilon/k_B$.
One can distinguish two stages: In the beginning, $H_4$ rises linearly
up to a saturation value.  This corresponds to a driven regime, where
first the instability described above
triggers the formation of a helix at the free chain end, and then this helix
propagates into the chain. In the second stage, the chain stays
perfectly helical for a while, until suddenly part of the helical order
gets lost. To analyze this second stage, we have recorded the lifetimes of
the helical state in 200 independent runs (different random numbers), starting
from an identical helical chain conformation. The lifetime was defined as the
time when the energy dropped below a given threshold $E_t$, which was chosen
below the initial energy $E_0$ such that $E_0 - E_t$ was of the order
$1 \epsilon$, much larger than the typical energy fluctuations of an intact
helix (roughly 0.002 $\epsilon$ per monomer). To ensure that the results did not
depend on the threshold, we carried out the analysis for at least two different
threshold values that differed by $1 \epsilon$, and checked that the results
did not change. This was done for a range of chain lengths, $N=25-100$,
and temperatures, $k_B T/\epsilon=0.04-0.045 $.
In all cases, the distribution of the lifetimes was nicely exponential. Two
examples are shown in Fig.~\ref{fig2}b). We conclude that the decay from the
``perfect'' helical state is a stochastic, rate-driven process. According to
Kramer's rate theory~\cite{risken}, the height of the energy barrier,
$\Delta E$, can be estimated from the temperature dependence of the escape
rate, $1/t_0 \propto \exp(\Delta E/k_B T)$. The fit of our data to this law
gives $\Delta E = 0.041 \epsilon$ (Fig.~\ref{fig2}c). Hence perfect helices
should persist for a long time at temperatures well below $\Delta E/k_B$,
and decay rapidly at temperatures above $\Delta E/k_B$.
This is indeed observed in the simulations.

\begin{figure}[t]
\includegraphics[scale=0.6,angle=0]{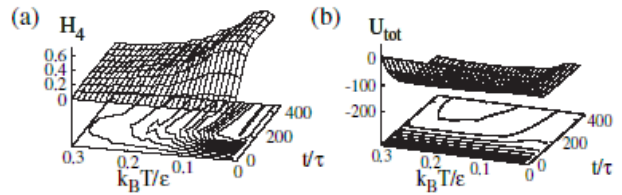}
\vspace*{-0.3cm}
\caption{\label{fig3}
Time evolution of the helical order parameter $H_4$ (a) and the total energy
$U_{tot}$ (b) in units of $\epsilon$ at temperatures between
$k_B T/\epsilon=0$ and $k_B T/\epsilon = 0.3$ for initially stretched chains,
averaged over 180 independent runs. The distance between levels in the
contour plots is 0.05 in (a) (lowest level 0.05)
and $20 \epsilon$ in (b)
(lowest level $-180 \epsilon$).
}
\end{figure}

More insight into the nature of the escape process can be gained from looking
at the dependence of the escape rate $1/t_0$ on the chain length $N$. We
assume that the escape is initiated by the nucleation of some defect.
If this defect is localized at the end of the chain, $1/t_0$ should not
depend on $N$; if it is localized somewhere in the middle, $1/t_0$ should
be proportional to $N$. Fig.~\ref{fig3}d) shows that the escape rate
increases with $N$, thus the escape defect forms in the middle of the
chain rather than at the end. However, the dependence is not strictly
linear; the true situation is more complicated than our simple argument
suggests.

At temperatures above $T=0.04 \epsilon/k_B$, the ``perfect'' helices decay
rapidly, but the system still retains helical order. To determine the
temperature range where long-lived helices appear, be they perfect or imperfect,
we have calculated the average  time evolution of $H_4$ for temperatures
up to $T=0.3 \epsilon/k_B$. The result is shown
in Fig.~\ref{fig3}a). Helices are observed at temperatures up to
$T \sim 0.1-0.15 \epsilon/k_B$. It is worth noting that in this
helical regime, the average total energy $U_{tot}$ is higher than at higher
temperatures ( Fig.~\ref{fig3}b)). This proves once more that the helical
state does not correspond to a true free energy minimum~\cite{fn2}.


Up to now, hydrodynamic effects mediated by the solvent
were disregarded. To assess the influence of the latter, we have also carried
out simulations of a dynamical model that includes hydrodynamic interactions
{\em via} an appropriate mobility tensor $D_{ij}$. Eq.~(\ref{eq:dynamics})
is then replaced by~\cite{ermak}
\begin{equation}
  \label{eq:hydrodynamics}
  \dot{\vec{r}}_i = \sum_j D_{ij} \vec{f}_j +
  \sum_j \xi_{ij} \vec{\eta}_j,
\end{equation}
where $\vec{\eta}$ is distributed as before (Eq.~(\ref{eq:eta})) and
$\xi$ fulfills $\xi \xi^T = D$. When determining the mobility tensor
for our chains, we must account for the fact that they are endgrafted:
The first bead is subject to a constraint force that ensures
$\vec{v}_1 \equiv 0$ in Eq.~(\ref{eq:hydrodynamics}). In the absence
of noise, the total force on bead 1 is thus given by
$\vec{f}_1 = - \sum_{j\ne 1}
[D_{11}^{\mbox{\tiny{free}}}]^{-1}
D_{1j}^{\mbox{\tiny{free}}}
\vec{f}_j$,
where $D_{ij}^{\mbox{\tiny{free}}}$ is the mobility tensor for free
chains. Hence the effective mobility matrix between the remaining beads is
\begin{equation}
\label{eq:mobility_grafted}
D_{ij} =
D_{ij}^{\mbox{\tiny{free}}} -
D_{i1}^{\mbox{\tiny{free}}}
[D_{11}^{\mbox{\tiny{free}}}]^{-1}
D_{1j}^{\mbox{\tiny{free}}}.
\end{equation}
The free mobility tensor $D^{\mbox{\tiny{free}}}$
was approximated by the Rotne-Prager
tensor~\cite{ermak,rotne} with solvent viscosity $\zeta$. The `square
root' tensor $\xi$ was determined by means of a Cholesky
decomposition of $D$.

\begin{figure}[t]
\includegraphics[scale=0.6,angle=0]{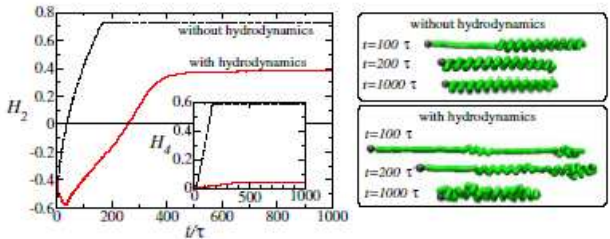}
\vspace*{-0.3cm}
\caption{\label{fig4}
Left:
Time evolution of the local helical order parameter $H_2$ (main plot) and the global
order parameter $H_4$ (inset) in an initially stretched chain of length
$N=50$ at temperature $k_B T=0.01 \epsilon$ without (dashed lines) and
with (straight lines) hydrodynamic interactions, averaged over 100 samples.
Right: Examples of corresponding conformations.
}
\end{figure}

Fig.~\ref{fig4} shows the average time evolution of the helical order in such a
chain at $T=0.01 \epsilon/k_B$. At this temperature, simple Langevin
motion yields highly ordered long-lived helices (cf. Fig.~\ref{fig1}).
In the presence of hydrodynamic interactions, the overall motion
of the chain is much more cooperative, and the local propagation mecanism
for helices does not work. As a result, the perfect global helical order
disappears, only local order remains.  We note in passing that the most
unstable chain mode, the zigzag mode, rises to much higher amplitudes and
persists much longer in the presence of hydrodynamics than without.

A more detailed account of the hydrodynamic simulations shall be given
elsewhere. Here, we just conclude that hydrodynamic interactions largely
reduce the effect reported above. The ``helical trap'' is avoided
in the presence of hydrodynamics. This is consistent with previous
simulations of chain collapse~\cite{yethiraj,kikuchi}, which also showed
that hydrodynamic interactions tend to inhibit trapping during
chain collapse.


To summarize, we have observed the spontaneous formation of well-ordered,
long-lived helices in Brownian dynamics simulations of simple flexible
model polymers at low temperatures. The effect relies on three conditions:
(i) The quality of the solvent must be sufficiently bad, {\em i.e.}, the
effective attractive interaction between monomers must be well above the
thermal energy $k_B T$. (ii) The chain must be straight initially. The
initial conformation then has unstable helical modes, which trigger
the helix formation. (iii) Hydrodynamic interactions should be screened.
This is the case, {\em e.g.}, in a porous or molecularly crowded environment.
Otherwise, the local ordering process is disturbed by global cooperative
processes. Remnant helicity (local helical order) is still observed in the
presence of hydrodynamic interactions, but the global order is much reduced.

We speculate that this effect might have contributed to establish helices
as one of the most important structural elements in biopolymers.
Recently, it has been hypothesized that the early molecular evolution of
life may have taken place in strongly confined environments, {\em e.g.},
thin interlinked mineral pores~\cite{baaske}. In such environments,
hydrodynamic interactions are largely screened, and chains collapsing from
stretched conformations may form nicely ordered helices. In an active system where
polymers constantly attach and detach to other objects and are permanently
stretched and released, helices might thus have appeared perpetually for purely
kinetic reasons.  They could then have been stabilized {\em a posteriori}
by suitable chemical modifications.


S.A.S. thanks the MHESR, Algeria, and the DFG, Germany (SFB 613) for
financial support during an extended visit to Bielefeld. The polymer
conformations have been visualized using
the open-source package VMD~\cite{vmd}.

\clearpage


\begin{thebibliography}{99}
\bibitem{fn0}
This is physically reasonable for large (bio)polymers in solution,
where the characteristic decay time of the drift velocity of the polymers
is usually much smaller (several orders of magnitude) than the
time scales characterizing the conformational dynamics.
\bibitem{fn1} Here, we have made use of the fact that the Lennard-Jones
potential is truncated at a distance $r_c < 3 b$. The general expression
is more complicated.
\bibitem{fn2} At thermodynamic equilibrium, the specific heat $dU/dT$
must be positive.
\end{thebibliography}
\end{document}